\documentclass[12pt,english]{article}

\usepackage{xcolor}

\makeatletter
\@ifpackageloaded{tex4ht}{
    \usepackage[dvips]{graphicx}
    \usepackage[tex4ht]{hyperref}
}{
    \usepackage{graphicx}
    \usepackage[colorlinks=true,urlcolor=blue]{hyperref}
}
\graphicspath{{images/}}

\usepackage[english]{babel}
\usepackage[utf8]{inputenc}
\usepackage{natbib}

\usepackage[margin=2.0cm,includehead]{geometry}

\usepackage{microtype}
\usepackage{xspace}

\usepackage{amsmath}
\usepackage{amssymb}
\usepackage{marvosym}

\usepackage{titling}

\usepackage{booktabs}
\usepackage{array}
\usepackage{tabularx}

\usepackage{cleveref}

\usepackage{setspace} 
\usepackage[nohyperlinks]{acronym}
\usepackage{textcase} 

\usepackage[switch]{lineno}
\usepackage{accsupp}

\usepackage{comment}

\usepackage[font=small,font=onehalfspacing,labelfont=bf]{caption}

\acrodef{AIF}{Arterial Input Function}
\acrodef{BCCA}{British Columbia Cancer Agency}
\acrodef{BLPOP}{Bilateral Parallel-Opposed Fields}
\acrodef{CT}{Computed Tomography}
\acrodef{CVLF}{Cross-Validation Lexicon Folding}
\acrodef{DCE-MRI}{Dynamic Contrast-Enhanced Magnetic Resonance Imaging}
\acrodef{DVH}{Dose-Volume Histogram}
\acrodef{IMRT}{Intensity-Modulated Radiotherapy Treatment}
\acrodef{KPS}{Karnofsky Performance Status}
\acrodef{LMS}{Least-Median of Squares}
\acrodef{LSS}{Least-Sum of Squares}
\acrodef{MRI}{Magnetic Resonance Imaging}
\acrodef{MR}{Magnetic Resonance}
\acrodef{NPLLR}{Non-Parametric Local Linear Regression}
\acrodef{NTCP}{Normal Tissue Complication Probability}
\acrodef{PTV}{Planning Target Volume}
\acrodef{QCD}{Quartile Coefficient of Dispersion}
\acrodef{QOL}[QoL]{Quality-of-Life}
\acrodef{QUANTEC}{Quantitative Analysis of Normal Tissue Effects in the Clinic}
\acrodef{REB}{Research Ethics Board}
\acrodef{RT}{Radiotherapy Treatment}
\acrodef{ROI}{Region of Interest}
\acrodefplural{ROI}[ROIs]{Regions of Interest}
\acrodef{UBC}{University of British Columbia}
\acrodef{VIF}{Venous Input Function}
\acrodef{fMRI}{Functional Magnetic Resonance Imaging}

\newcommand{\eg}{e.g.,\ }       
\newcommand{\ie}{i.e.,\ }       
\newcommand{\vs}{vs.\ }

\newcommand{\etc}{etc.\xspace}

\newcommand{\cf}{cf.\xspace}    
\newcommand{\nb}{n.b.\xspace}

\newcommand{\DICOMautomaton}{{\small \texttt{DICOM\-auto\-maton}}\xspace}

\hypersetup{
    bookmarks=false,         
    unicode=false,           
    pdftoolbar=false,
    pdfmenubar=false,
    pdffitwindow=false,
    pdfstartview={FitW},
    pdfauthor={Haley Clark},
    pdfcreator={Haley Clark},
    pdfproducer={Haley Clark},
    pdfnewwindow=true,       
    colorlinks=true,         
    linkcolor=blue,          
    citecolor=blue,          
    urlcolor=blue
}

\excludecomment{trimmedforpublication}

\begin{document}

\newgeometry{margin=2.5cm}

\newcommand{\titleaux}[1]{\gdef\puB{#1}}
\newcommand{\puB}{}
\renewcommand{\maketitlehookc}{\vspace{1cm}\par\noindent \puB\vspace{1cm}}

\title{Prefer Nested Segmentation to Compound Segmentation
       \vfill}

\author{
  \large 
  Haley D.\ Clark\textsuperscript{\textdagger,1,2},\,\, 
  Stefan A.\ Reinsberg\textsuperscript{1},\,\, 
  Vitali Moiseenko\textsuperscript{3},\,\, \\
  \large 
  Jonn Wu\textsuperscript{1,4},\,\, and 
  Steven D.\ Thomas\textsuperscript{2}. 
  \and 
  \small 
  \textsuperscript{1}Department of Physics and Astronomy, \\
  \small 
  University of British Columbia, \\
  \small 
  6224 Agricultural Road, \\
  \small 
  Vancouver, BC, V6T 1Z1, Canada. \\
  \and 
  \small 
  \textsuperscript{2}Department of Medical Physics, \\
  \small 
  British Columbia Cancer Agency, \\
  \small 
  600 West Broadway, \\
  \small 
  Vancouver, BC, V5Z 4E6, Canada. \\
  \and 
  \small 
  \textsuperscript{3}Department of Radiation Medicine and Applied Sciences, \\
  \small 
  University of California -- San Diego, \\
  \small 
  9500 Gilman Drive, \\
  \small 
  La Jolla, Ca, 92093, USA. \\
  \and 
  \small 
  \textsuperscript{4}Department of Medicine, \\
  \small 
  University of British Columbia, \\
  \small 
  2775 Laurel Street, 10\textsuperscript{th} Floor, \\
  \small 
  Vancouver, BC, V5Z 1M9, Canada. \\
}

\titleaux{\vfill
  \begin{itemize}
    \item[\textdagger] Corresponding author: 
      H.\ D.\ Clark, via \url{http://www.halclark.ca/Contact.html}. 
    \item Disclaimer: the views expressed in this manuscript are our own and are not the official position of our employers or
      funders.
  \end{itemize}
  \vspace{1cm}
}

\date{
      \today}

\maketitle
\normalsize

\restoregeometry

\newpage
\nolinenumbers
\onehalfspacing

\begin{abstract}
\noindent
\textbf{\small Introduction:} 
  Intra-organ radiation dose sensitivity is becoming increasingly relevant in clinical radiotherapy. One method for
  assessment involves partitioning delineated regions of interest and comparing the relative contributions or importance
  to clinical outcomes. We show that an intuitive method for dividing organ contours, compound (sub-)segmentation, can
  unintentionally lead to sub-segments with inconsistent volumes, which will bias sub-segment relative importance
  assessment. An improved technique, nested segmentation, is introduced and compared.
\newline
\noindent
\textbf{\small Materials and Methods:} 
  Clinical radiotherapy planning parotid contours from 510 patients were segmented. Counts of radiotherapy dose matrix
  voxels interior to sub-segments were used to determine the equivalency of sub-segment volumes. The distribution of
  voxel counts within sub-segments were compared using Kolmogorov-Smirnov tests and characterized by their dispersion.
  Analytical solutions for two- and three-dimensional analogues were derived and sub-segment area/volume were compared
  directly.
\newline
\noindent
\textbf{\small Results:} 
  Both parotid and 2D/3D region of interest analogue segmentation confirmed compound segmentation 
  intrinsically produces sub-segments with volumes that depend on the region of interest shape and
  selection location. 
  Significant volume differences were observed when
  sub-segmenting parotid contours into $18$\textsuperscript{ths}, and vanishingly small sub-segments were observed when
  sub-segmenting into $96$\textsuperscript{ths}. Central sub-segments were considerably smaller than sub-segments on the
  periphery. Nested segmentation did not exhibit these shortcomings and produced sub-segments with equivalent volumes
  when dose grid and contour collinearity was addressed, even when dividing the parotid into $96$\textsuperscript{ths}.
  Nested segmentation was always faster or equivalent in runtime to compound segmentation.
\newline
\noindent
\textbf{\small Conclusions:} 
  Nested segmentation is more suited than compound segmentation for analyses requiring equal weighting of sub-segments.
\end{abstract}


\newpage
\nolinenumbers
\onehalfspacing

\section*{Introduction}

Heterogeneous functional dose response for organs-at-risk is becoming increasingly relevant for clinical radiotherapy
planning. In 2005, \citet{konings2005volume} found evidence of region-dependent volume effects in rat
parotid. Years later in 2010, as part of the encompassing \emph{Quantitative Analysis of Normal Tissue Effects in the
Clinic} (QUANTEC) organ-focused reviews for clinical guidelines, \citet{deasy2010radiotherapy} concluded that
better predictive models were needed to model xerostomia risk. One factor recommended for investigation was whether
regions within the parotid could be located that exhibited variable dose sensitivity, increased or decreased functional
burden, or otherwise controlled function preservation to a higher degree than surrounding tissues. Other articles in the
same report provided similar recommendations for other organs
\citep{rancati2010radiation,dawson2010radiation,pan2010radiation}. In response, organ \acp{ROI} are increasingly being
segmented or handled heterogeneously to model dose response to various aspects within the organ. Reports of trials
underway are emerging \citep{vanluijk2015sparing,miah2016recovery,xiao2016split}.

Methods for complex contour segmentation, including planar segmentation, have been described in the literature
\citep{clark2014automated}. A recent paper by \citet{vanluijk2015sparing} made use of a planar
segmentation method in which fractional volumes were used to implement a compounded Boolean sub-segment selection
mechanism. Here we show that, perhaps unintuitively, such a scheme will result in sub-segments of differing volume
depending on the shape of the \ac{ROI} and selection location within it. Inconsistent segmentation volumes can be
problematic for investigation of sub-organ effects because sub-segments will represent inconsistent portions of the
whole \ac{ROI}. Performing sensitivity analysis, model fitting, or tests of associativity (\eg correlation) will result
in bolstered or undermined sub-segment importance, model parameters, or associativity, which must be corrected.

We propose an improvement, which we call \emph{nested segmentation}, that is ``fair'' in the sense that it will produce
equal-volume sub-segments uniformly throughout the \ac{ROI} when the cleaving method is free of bias. Furthermore, it is
robust -- if the cleaving method \emph{is} biased, as-fair-as-possible sub-segments are produced. It is also faster than
compound segmentation, requiring equivalent or less geometrical processing. An implementation based on segmentation of
planar contours is tested using clinical data from 510 head-and-neck cancer patients.

We also present two methods that can be used in conjunction with segmentation that help ensure an equal number of grid
voxels (\eg radiotherapy dose matrix voxels) are contained within the boundaries of each sub-segment: oblique cleaving
planes and grid supersampling. We show that both methods ameliorate issues arising from collinearity of dose grid and
\ac{ROI} boundaries.

\section*{Materials and Methods}
\subsection*{Segmentation}

Segmentation\footnote{Alternatively \emph{contour sub-segmentation} or just \emph{sub-segmentation}, to differentiate it
from \emph{image} segmentation.} refers to the process in which part of a volume delineated by closed contour lines (\ie
a \ac{ROI}) is partitioned into connected pieces (``sub-segments'') and one or more are retained (``selected''). We
refer specifically to segmentation, but the process is equivalent to \emph{volume truncation} for polyhedra and generic
\emph{division} or \emph{partitioning} of areas, volumes, and hypervolumes (\eg geometric primitives, such as triangles,
spheres, and cubes). Sub-segment selection can be accomplished in a variety of ways, but in this work we focus on the
method described by \citet{vanluijk2015sparing}, which we refer to as \emph{compound} segmentation. Our
improved method, nested segmentation, is believed to be more robust toward the `fair distribution' problems of ensuring
that selected sub-segments have equivalent volume and contain an equivalent number of entities (\eg dosimetric grid
voxels) regardless of the \ac{ROI} shape and selection location (\eg periphery \vs centre). Salient differences are
described in the following subsections.

\subsection*{Compound Segmentation}

\begin{figure*}[htbp]
  \centering
  \includegraphics[width=0.95\linewidth]{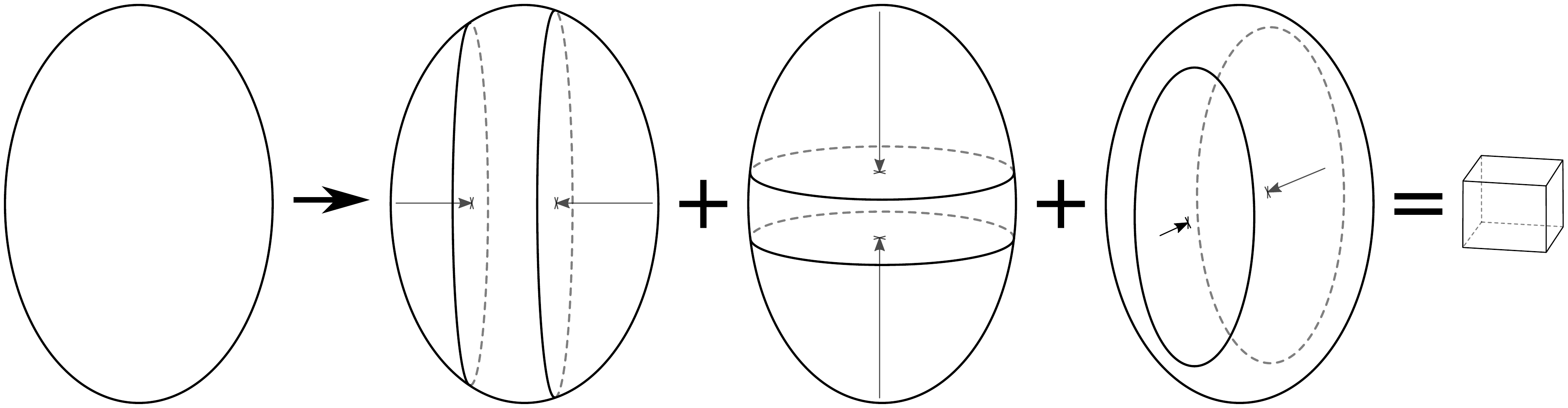}
  \caption{Demonstration of compound segmentation with three parallel pairs of mutually orthogonal planes (six
           planes in total). 
           \label{fig:vanLuijkSubSegDemo}}
\end{figure*}

Compound segmentation is a planar segmentation technique that makes use of (infinite) cleaving planes. The cleaved
sub-segment faces are flat, and thus when the \ac{ROI} is convex all cleaved surfaces remain convex. Compound
segmentation proceeds by specification of cleaving plane orientations and volume percentiles (\ie fractional volumes).
Each fractional volume ($f \in [0,1]$) unambiguously specifies a cleaving plane which contains $f$ on one side of the
plane and $1-f$ on the other\footnote{Both $f=0$ and $f=1$ are ambiguous because they are not unique. The ambiguity is
not relevant for segmentation.}. Each plane requires a single unit vector or two free parameters to orient the plane. In
\citep{vanluijk2015sparing} six planes are located and used to select a sub-segment interior to the boundary of a parotid
\ac{ROI}. There are three sets of parallel planes; each set is orthogonal to the others (see figure
\ref{fig:vanLuijkSubSegDemo}). Use of one less plane would permit selection of an arbitrary sub-segment with a single
portion of the \ac{ROI} surface\footnote{In the case of convex \acp{ROI}.}, use of two less planes would permit either
one or two disjoint \ac{ROI} surface portions, \etc.

In compound segmentation, all cleaving planes are derived using percentiles or fractional volumes that refer to the whole
\ac{ROI}. Only after all planes are located is segmentation performed by application of cleaving planes
to the \ac{ROI} volume, and only the interior is selected.

\subsection*{Nested Segmentation}

\begin{figure*}[htbp]
  \centering
  \includegraphics[width=0.95\linewidth]{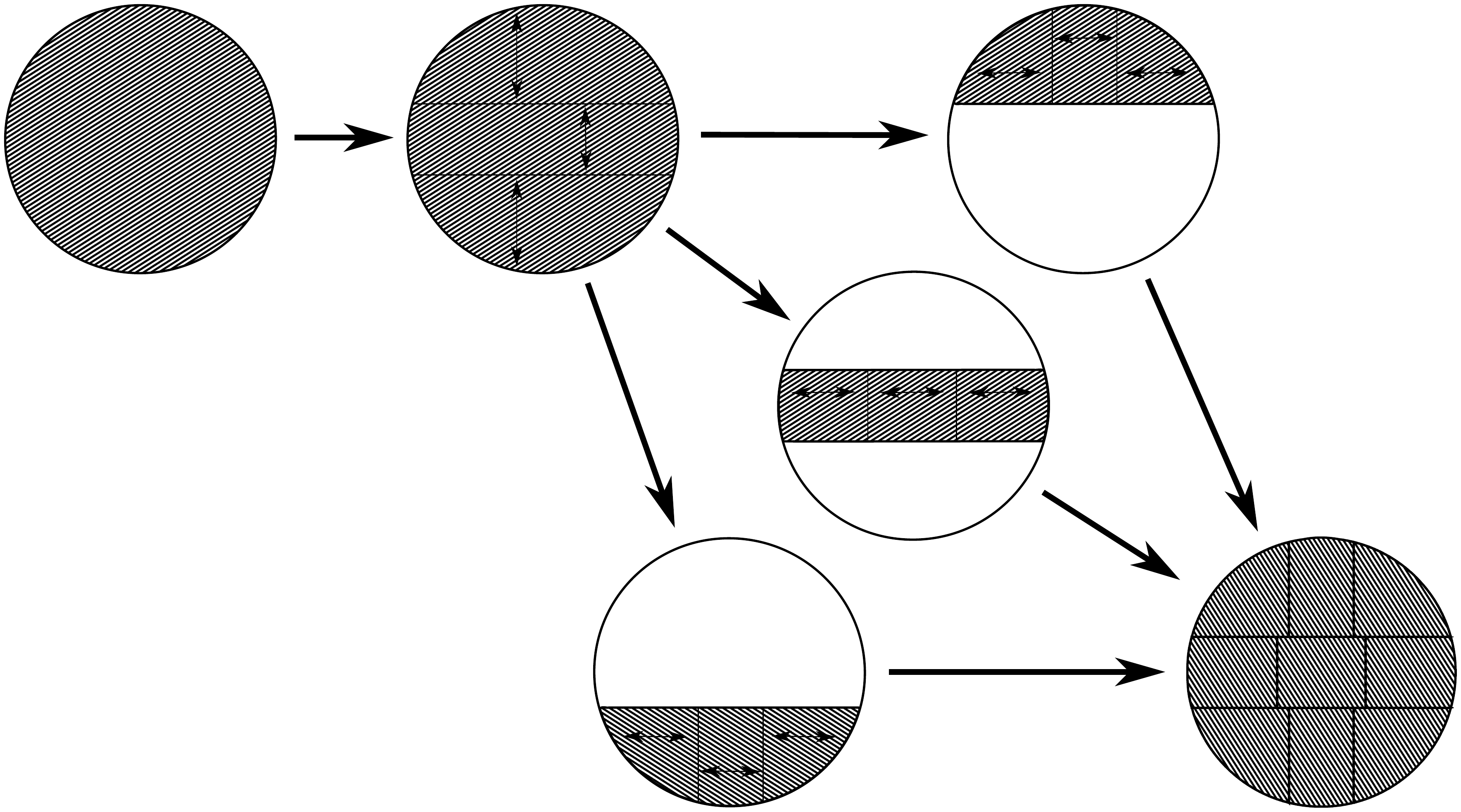}
  \caption{Demonstration of nested segmentation on a circle with $f = 1/3$. 
           All sub-segments have area $\pi r^{2} / 9$.
           \label{fig:NestedSubsegGist}}
\end{figure*}

We propose an improved method in which sub-segments are selected as the interior region between two parallel planes, as
with compound segmentation, however cleaves are performed eagerly, before the next pair of planes can be located. The
location of cleaving planes are thus derived from the volume of \emph{remaining} sub-segments, not the original \ac{ROI}
(see \cref{fig:NestedSubsegGist}). As each individual stage of segmentation achieves a fair divvy of the remaining
volume, sub-segments are expected to always contain an equivalent portion of the \ac{ROI} volume when the partitioning
method is fair.

\subsection*{\acs{ROI} Segmentation Comparison}

We compare segmentation methods on a data set of 510 parotid gland \acp{ROI} from 510 head-and-neck cancer patients (one
per patient, to avoid any potential bias from shape correlation between left and right parotid). We perform segmentation
to generate $3$, $18$, and $96$ sub-segments from each \ac{ROI}. $18$\textsuperscript{ths} and $96$\textsuperscript{ths}
segmentation used $6$ cleaving planes (\ie three mutually orthogonal sets of parallel planes) for each sub-segment
whereas segmentation into thirds used a single pair of planes parallel to the \ac{ROI} contours. Because \acp{ROI} are
defined in terms of equidistant, parallel, planar contours with no gaps, contour area is used as a surrogate for volume.

Each parotid in this data set includes a radiotherapy treatment planning dosimetric grid which is used to derive dose-volume
statistics of interest. We compute the number of voxels lying within each sub-segment and compare distributions for each
sub-segment. Cardinal axes-aligned cleaving planes are often desired for ease of specification or bounding of anatomical
regions (\eg posterior region, lateral-caudal region, etc.). However, raster grids are commonly aligned with the
cardinal directions, and are in this data set, which makes perfectly fair partitioning of voxels impossible (\ie due to
collinearity; a row locally aligned with the contour boundary is either within or outside of the sub-segment, but the row
may contain many voxels). We employ two techniques which can help to more fairly partition sub-segments: raster grid
supersampling and oblique cleaving planes. Supersampling used fine ($15\times$) cubic interpolation so that each voxel
was effectively interpolated into $225$ voxels. A cyclic rotation of $22.5^{\circ}$ between cardinal axes was used to
orient oblique planes.

Contours are operated on directly rather than rasterizing them onto a volumetric grid. Bisection is used to locate
planes corresponding to the requisite $f$. The bisection method used for $\mathbb{R}^3$ \ac{ROI} segmentation had a
stopping tolerance set to $1\%$ for all clinical data segmentation, but contours lying in planes parallel to the
cleaving plane were treated atomically and were thus indivisible. Parotids with few contours therefore could not achieve
$1\%$ tolerance. Voxels for each dose matrix were of fixed volume, so summary statistics about the distribution of
sub-segment voxel counts estimate sub-segment volume.

Segmentation into thirds employed only two planes parallel to contours. Nested and compound
segmentation should produce identical results in this case because only a single pair of cleaves are performed. However,
it is challenging because contours are not divided in this case. We include the comparison to demonstrate that bisection
is produces sufficiently fair sub-segments. All \ac{ROI} and dose manipulations were performed using
\DICOMautomaton \citep{clark2014automated,clark2014contour}.

\subsection*{Statistics}
Distributions of volumes and voxels counts for sub-segments were compared using a non-parametric Kolmogorov-Smirnov
test. The null hypothesis is that one of the distributions is drawn from the same parent distribution as the other, so
that the distributions are statistically identical. Individual sub-segment volumes are compared with other sub-segments
in the same \ac{ROI} using voxel counts to determine the spread due to the cumulative effects of the segmentation method
and raster grid voxel alignment. The \ac{QCD} provides a normalized measure of the dispersion of sub-segment areas for
each patient \citep{bonett2006confidence}. Median \ac{QCD} and median-normalized ranges are reported to characterize the
population. A standard statistical significance threshold ($\alpha$) of 0.05 was used.

\section*{Results}

\subsection*{Analytic Comparison}

\begin{figure}[htbp]
  \centering
  \includegraphics[width=0.90\linewidth]{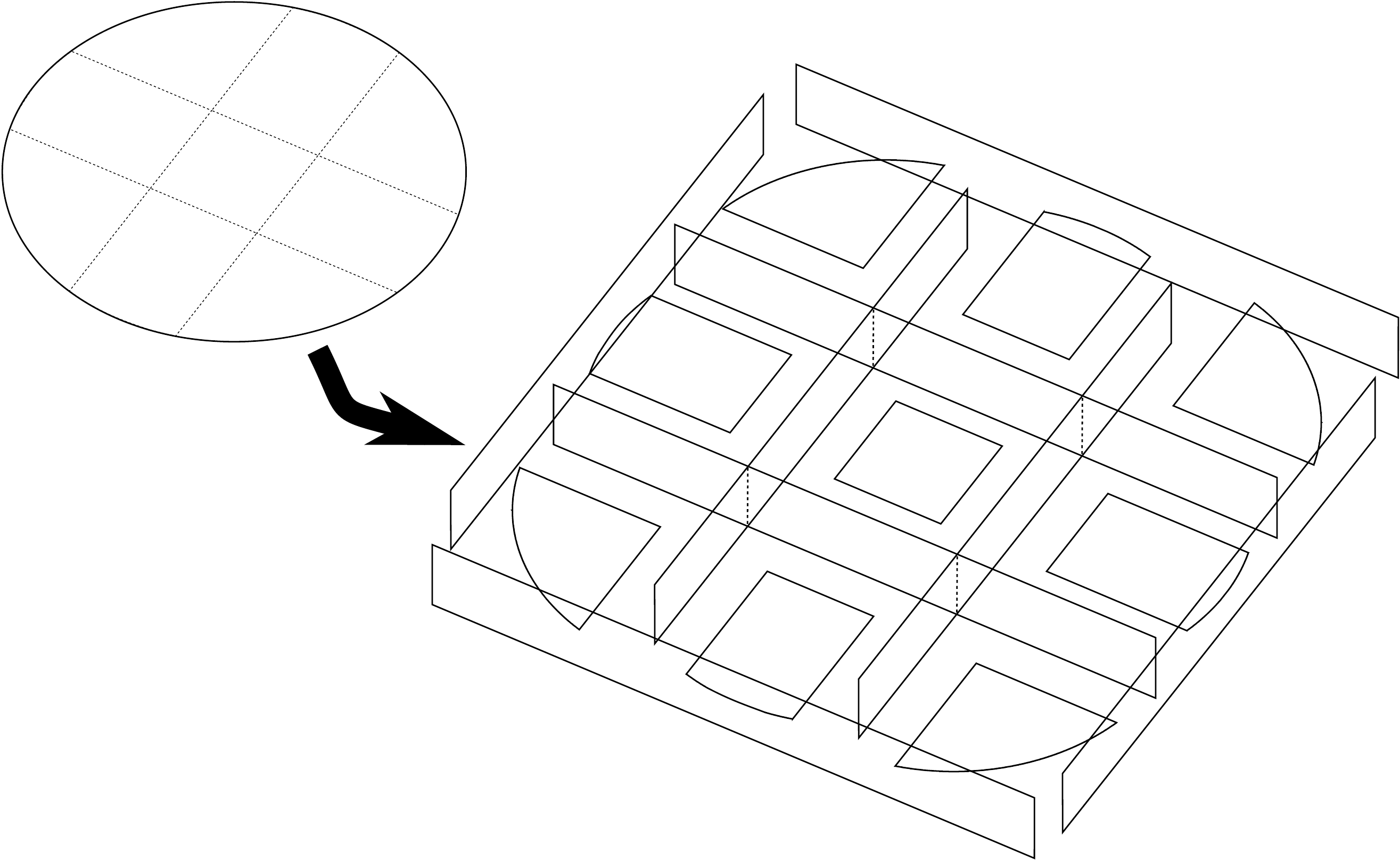}
  \caption{Partitioning of circle into nine sub-segments using compounded segmentation (exploded view).
           Each sub-segment is bounded by two parallel pairs of mutually orthogonal planes.
           \label{fig:PartitioningCircleCompoundedPlanes}}
\end{figure}

\begin{figure}[htbp]
  \centering
  \includegraphics[width=0.50\linewidth]{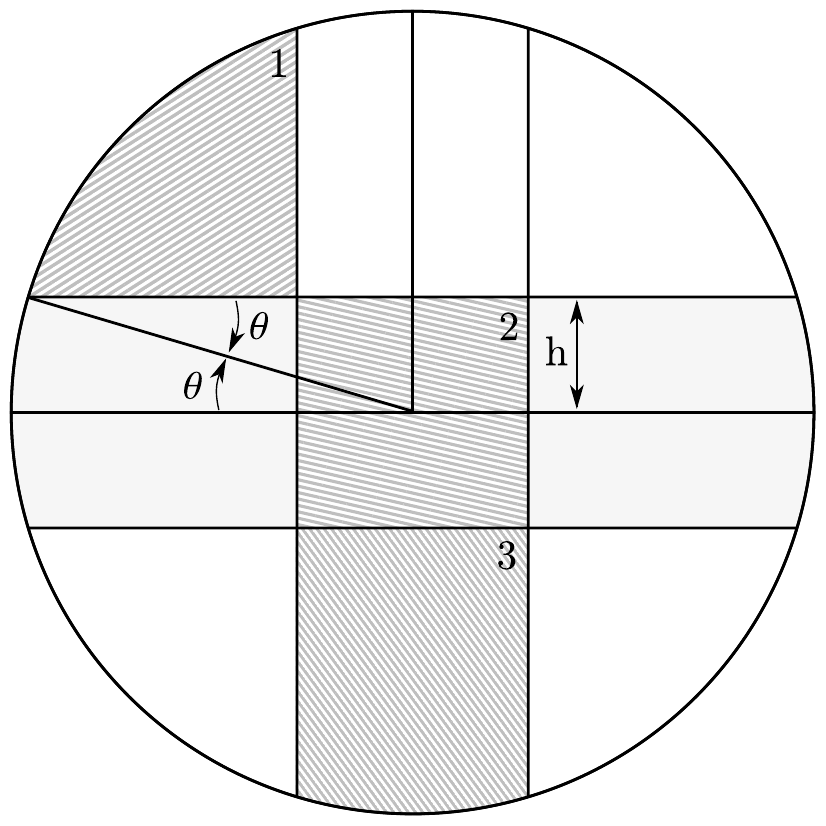}
  \caption{Calculation of sub-segment areas in terms of area of a wedge, right triangle, and square 
           defined by $f = 1/3$, $h \approx 0.264932r$, and $r$.
           The three distinct types of sub-segments are shown: (1) ``corner,'' (2) ``centre,'' and (3) ``centre-adjacent.''
           \label{fig:CircleAreaDerivationPrimitives}}
\end{figure}

The $\mathbb{R}^2$ analog to \ac{ROI} segmentation is individual planar contour segmentation. We segment a circle of
radius $r$ into nine sub-segments using compound segmentation (see \cref{fig:PartitioningCircleCompoundedPlanes}). The
fractional area on the small side of each cleaving plane is $1/3$.
Cleaving plane orientations are fixed, but the offsets from the origin are unknown and are derived analytically or
located through, \eg bisection. Using elementary methods,
it can be shown that the fractional area enclosed by a plane offset from the origin (\nb a secant line) and a parallel
plane intersecting the origin, in terms of their separation ($h \in [0,r]$; \ie the \emph{apothem}; \cf
\citep{archibald2015euclid}) is
\begin{align}
    \label{eqn:FractionalAreaHalfCircle}
    f = \frac{2}{\pi} \left( \arcsin{\frac{h}{r}} + \frac{h^2}{r^2} \cot \arcsin{ \frac{h}{r} } \right).    
\end{align}
Inversion is used to determine $h$. When $f=1/3$, $h \approx 0.264932r$. Derivation of the nine sub-segment areas is
then straightforward (see \cref{fig:CircleAreaDerivationPrimitives}). Results are summarized in
\cref{tab:CompoundSubSegFractionalAreas}. The smallest, as a ratio of the `fairly distributed' area ($\pi r^{2} / 9$) is
the centre sub-segment at $\approx 0.8043$; the centre-adjacent sub-segments are the largest at $\approx 1.0978$. 

\begin{table}[ht]
\centering
\begin{tabular}{lll}
  \toprule
                       & \multicolumn{2}{c}{Ratio of Fair} \\
    \cmidrule{2-3}
 Sub-segment           & (general $f$)  & ($f=1/3$)   \\
  \midrule
 centre                & $\frac{36}{\pi}\left(\frac{h}{r}\right)^{2}$ 
                       & $0.804306\times$ \\
 centre-adjacent       & $\frac{9}{2} f - \frac{18}{\pi} \left(\frac{h}{r}\right)^{2}$ 
                       & $1.097847\times$ \\
 corner                & $\frac{9}{4}\left(1-2f\right) + \frac{9}{\pi} \left(\frac{h}{r}\right)^{2}$ 
                       & $0.951077\times$ \\
  \bottomrule
\end{tabular}
  \caption{Ratios of the fair fractional area for compound segmentation sub-segments in terms of the apothem ($h$) and 
           fractional area ($f$). All ratios are fractions
           of the fairly distributed area ($\pi r^{2} / 9$) in which each sub-segment has an equivalent area.
           Centre sub-segments have four planar edges, centre-adjacent
           have three, and corner sub-segments have two.
           \label{tab:CompoundSubSegFractionalAreas}}
\end{table}

\begin{figure}[htbp]
  \centering
  \includegraphics[width=0.75\linewidth]{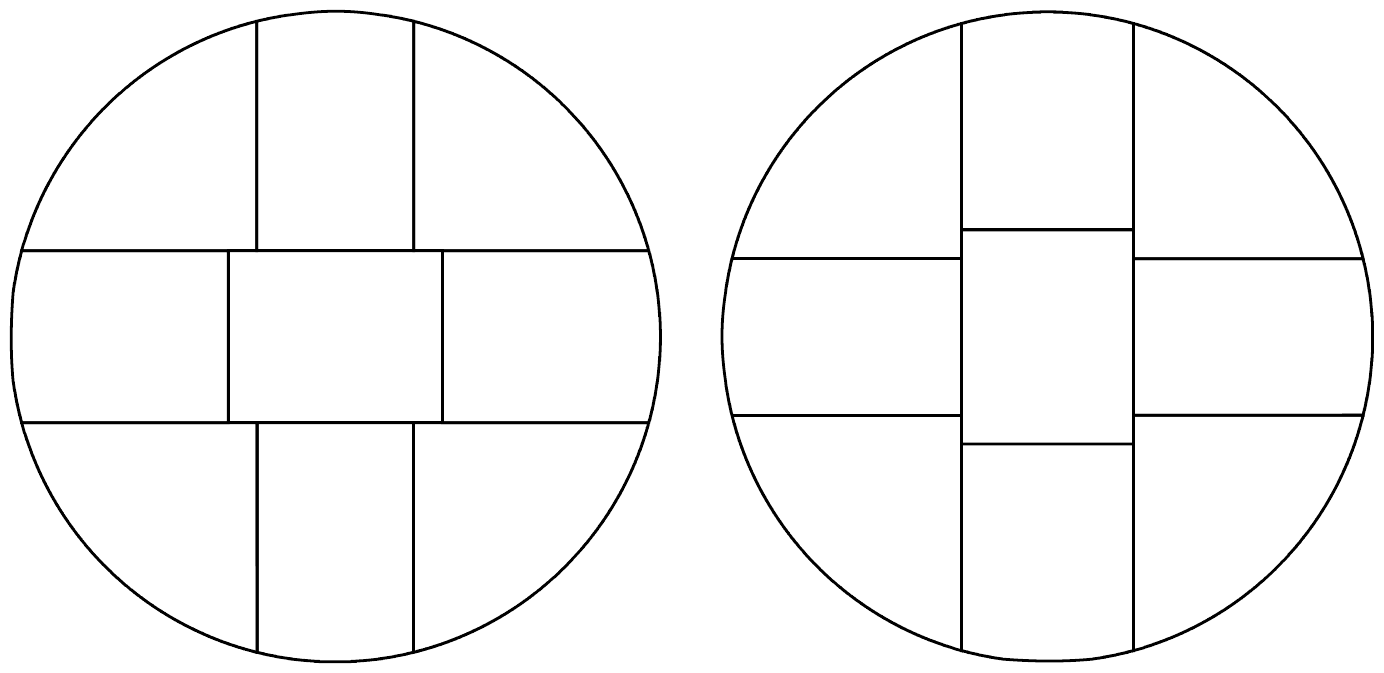}
  \caption{Nested segmentation of a circle into nine sub-segments each with area $\pi r^{2}/9$.
           The orientation of the first cleave can be chosen two ways. Both are shown.
           The cleaving order is important in nested segmentation but not for compound segmentation.
           \label{fig:NestedSubsegCircleTwoWays}}
\end{figure}

Nested segmentation, on the other hand, generated sub-segments with equal area (see
\cref{fig:NestedSubsegCircleTwoWays}). If all partitions can be made fairly, so that a cleaving plane that achieves the
desired fractional areas is located exactly, then each sub-segment area is tautologically known as the product of
requested fractional areas. For example, each final sub-segment in figure \ref{fig:NestedSubsegGist} has an area $1/3$
of $1/3$ of the total. The first cleave is identical to compound segmentation and so the apothem is given by
\cref{eqn:FractionalAreaHalfCircle}. Because nested segmentation is a greedy algorithm and the first cleave does not
take into account later cleaves, sub-segments are in general asymmetric. The two asymmetries possible for segmentation
into nine sub-segments (\nb with fixed cleave plane orientations) are shown in \cref{fig:NestedSubsegCircleTwoWays}.

Moving to $\mathbb{R}^{3}$, compound segmentation applied to a sphere partitioned into $3 \times 3 \times 3 = 27$
sub-segments yielded a centre sub-segment volume $\approx 0.596$ that of the fair volume. Centre-adjacent-adjacent
sub-segments had areas $\approx 1.105$ that of the fair volume. Nested segmentation again produced fair volumes that were
tautologically, in this case, $1/27$\textsuperscript{th} of the whole. A sphere constructed of discrete stacks of contours
sharing a planar orientation, which is common for medical image \acp{ROI}, approached both compound and nested
segmentation results asymptotically as the contour thickness shrunk.

\subsection*{Segmentation into Thirds}

Using compound segmentation, whole \ac{ROI} were segmented into three sub-segments ($f_{\rm{axial}}$ spanned
$\left[0,\frac{1}{3}\right]$, $\left[\frac{1}{3},\frac{2}{3}\right]$, and $\left[\frac{2}{3},1\right]$). Bisection was
employed and cleaving planes were held parallel to contours. The median number of voxels in each sub-segment spanned
$587.5-605.0$. The distribution of voxel counts in cranial, middle, and caudal sub-segments were compared with a
Kolmogorov-Smirnov test. Each unique comparison in \{left, right\} parotid $\otimes$ \{cranial, middle, caudal\}
sub-segments was performed, yielding ten tests. In all cases the two-sided $p > 0.20$. These tests indicate the
bisection approach results in appropriately partitioned sub-segments that contain $1/3$ of the original parotid volume
without systematic bias detectable at the $\alpha = 0.05$ level. Results were identical for nested segmentation.

\subsection*{Segmentation into 18\textsuperscript{ths}}

\begin{figure*}[htb] 
  \centering
  \includegraphics[width=0.95\linewidth]{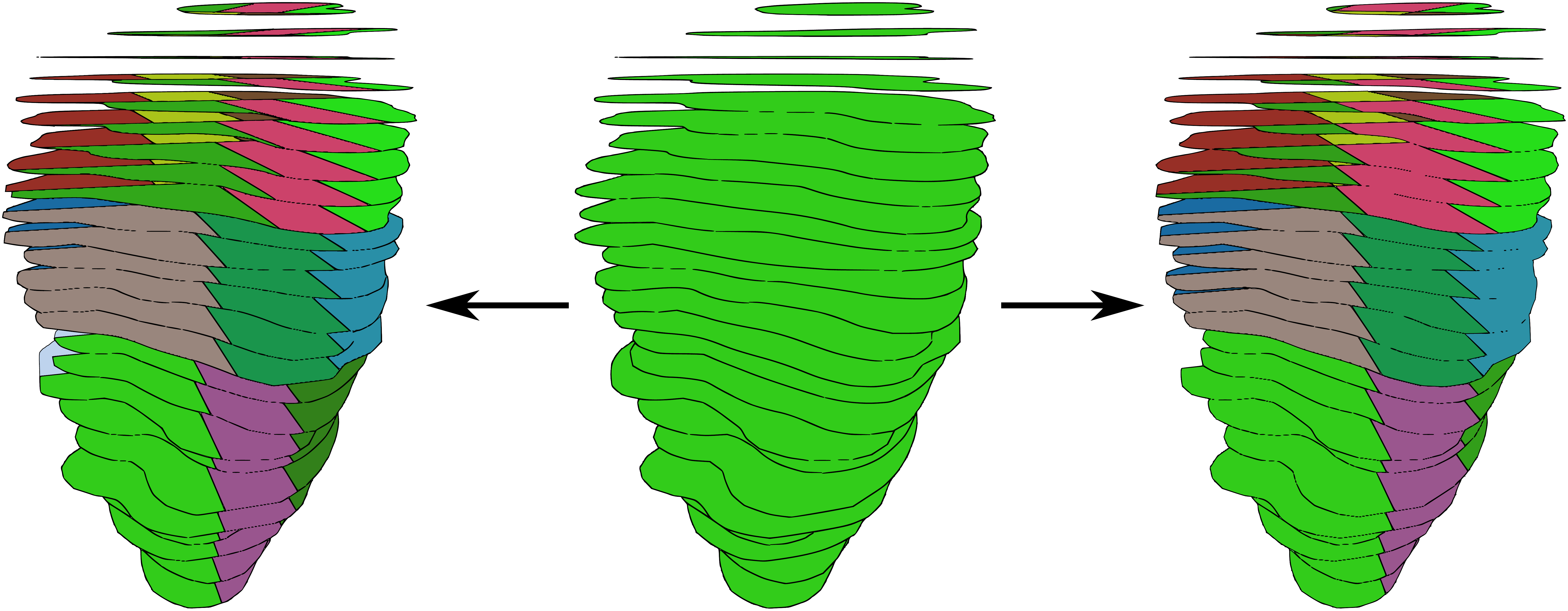}
  \caption{Depiction of nested (left) and compound (right) segmentation of whole parotid (centre) into $18$ sub-segments.
           \label{fig:DepictionOfWholeTo18thsSubseg}}
\end{figure*}

Figure \ref{fig:DepictionOfWholeTo18thsSubseg} shows nested and compound segmentation of whole parotid into $18$
sub-segments. Sub-segments are composed of axially-adjacent slices coloured uniformly\footnote{Colours were chosen for
maximum contrast using a modification of the palette described in \citep{kelly1965twenty}.}. Both $f_{\rm{axial}}$ and
$f_{\rm{sagittal}}$ spanned $\left[0,\frac{1}{3}\right]$, $\left[\frac{1}{3},\frac{2}{3}\right]$, and
$\left[\frac{2}{3},1\right]$; $f_{\rm{coronal}}$ spanned $\left[0,\frac{1}{2}\right]$ and $\left[\frac{1}{2},1\right]$.
The cleaving order was axial $\rightarrow$ coronal $\rightarrow$ sagittal. As can be seen in figure
\ref{fig:DepictionOfWholeTo18thsSubseg}, nested and compound method sub-segment locations differ only slightly.
However, it is apparent that sub-segments in the compound method do not all have equivalent volume.

Using compound segmentation without supersampling or oblique cleaving planes, sub-segment voxel counts had a mean of
$100.0$ voxels within each sub-segment (std.\ dev.\ $=64.0$; std.\ dev.\ of the mean $=0.7$; median $= 92.0$). The
median number of voxels in each sub-segment spanned $38-152$. Only $57.6\%$ of sub-segment voxel counts had an absolute
percent difference of less than $50\%$ of the mean. Direct comparison of the voxel count distributions within
sub-segments was performed using Kolmogorov-Smirnov tests. Unique comparison of all $18$ sub-segments required $153$
tests -- in $124$ cases ($81\%$) the null hypothesis failed to be rejected and distributions were found to differ
significantly (\ie $p < 0.05$ in $124$ cases). Skewness of the combined voxel count distribution was 
$0.991$ using the ratio of moments technique, which indicates a strong positive skew. However, the mean voxel count in
each \emph{type} of sub-segment were more symmetrically distributed with a skewness of $0.075$ and a std.\ dev.\ $=
28.0$. No significant correlation was detected between the average sub-segment mean voxel count and position relative to
the parotid centre (\eg with relativity denoted by $-1$, $0$, or $+1$ in the cardinal directions).

Using nested segmentation without supersampling or oblique cleaving planes, sub-segment voxel counts again had a mean of
$100.0$ voxels within each sub-segment (std.\ dev.\ $=49.8$; std.\ dev.\ of the mean $=0.6$; median $=97.0$). However
the median number of voxels in each sub-segment spanned $92.5-101$. $70.2\%$ of sub-segment voxel counts had an absolute
percent difference of less than $50\%$ of the mean. Direct comparison of the voxel count distributions within
sub-segments using Kolmogorov-Smirnov tests showed that the null hypothesis failed to be rejected in only $2$ of $153$
($1.3\%$) cases (\ie $p < 0.05$ in $2$ cases).

\subsection*{Segmentation into 96\textsuperscript{ths}}

\begin{figure*}[htb] 
  \centering
  \includegraphics[width=0.95\linewidth]{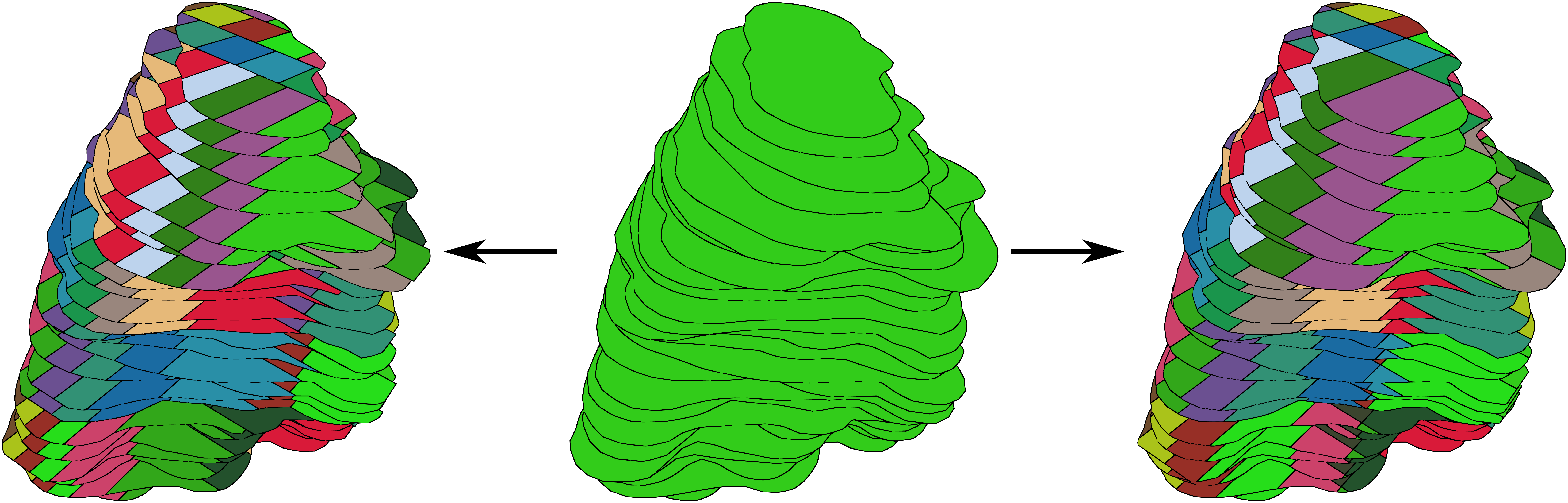}
  \caption{Depiction of nested (left) and compound (right) segmentation of whole parotid (centre) into $96$ sub-segments.
           \label{fig:DepictionOfWholeTo96thsSubseg}}
\end{figure*}

For segmentation into $96$\textsuperscript{ths}, both $f_{\rm{axial}}$ and $f_{\rm{coronal}}$ spanned
$\left[0,\frac{1}{4}\right]$, $\left[\frac{1}{4},\frac{1}{2}\right]$, $\left[\frac{1}{2},\frac{3}{4}\right]$, and
$\left[\frac{3}{4},1\right]$ whereas $f_{\rm{sagittal}}$ spanned $\left[0,\frac{1}{6}\right]$,
$\left[\frac{1}{6},\frac{1}{3}\right]$, $\left[\frac{1}{3},\frac{1}{2}\right]$, $\left[\frac{1}{2},\frac{2}{3}\right]$,
$\left[\frac{2}{3},\frac{5}{6}\right]$, and $\left[\frac{5}{6},1\right]$. The cleaving order was axial $\rightarrow$
coronal $\rightarrow$ sagittal. As can be seen in figure \ref{fig:DepictionOfWholeTo96thsSubseg}, nested and compound
segmentation again produce similar-shaped sub-segments in roughly similar locations. However, compound segmentation
produces sub-segments with substantially different volumes, such as those with vanishingly small volume (in the
centre-bottom of \cref{fig:DepictionOfWholeTo96thsSubseg}; right side). The comparable nested segmentation sub-segments,
on the other hand, are larger and have the same apparent volume as all other sub-segments (left side of
\cref{fig:DepictionOfWholeTo96thsSubseg}). Nested method sub-segment median \ac{QCD} were less disperse than the
compound method ($0.097$ \vs $0.37$). Oblique planes reduced dispersion nearly $25\times$ for nested method
sub-segments ($0.097 \rightarrow 0.0041$). Conversely, they \emph{increased} compound method dispersion ($0.37
\rightarrow 0.46$).

\begin{table*}[ht]
\centering
\resizebox{\linewidth}{!}{
\begin{tabular}{lccccc}
  \toprule
                       & \multicolumn{1}{c}{Compound} & \multicolumn{4}{c}{Nested} \\
  \cmidrule(lr){2-2} \cmidrule(lr){3-6}                       
                       & Oblique + Supersampling      & Unmodified       & Oblique    & Supersampling  & Oblique + Supersampling  \\
  \midrule
   Median range        & $91.5-13829$ & $20-30$  & $18-20$ & $4027.5-4200$ & $4212-4287$ \\
   Median              & $5558$ & $24$ & $19$ & $4103$ & $4246$ \\
   Range/Median        & $2.47$ & $0.417$ & $0.105$ & $0.042$ & $0.018$ \\
   Sig.\ K-S tests    & $2435$ $(53.3\%)$ & $258$ $(5.7\%)$ & $1$ $(0.02\%)$ & $0$ $(0\%)$ & $0$ $(0\%)$ \\
   \ac{QCD}           & $0.46$ & $0.097$ & $0.0041$ & $0.097$ & $0.0041$ \\
   Runtime             & $143\,ms$ & $36\,ms$ & $29\,ms$ & $135\,ms$ & $131\,ms$ \\
  \bottomrule
\end{tabular}
}
  \caption{Comparison of median voxel counts, quartile coefficients of dispersion (QCD), and runtime for compound and nested segmentation.
           Sig.\ K-S tests refers to the number of statistically significant Kolmogorov-Smirnov tests (out of $4560$; $\alpha =
           0.05$). Runtime is per (individual) sub-segment and was measured on an Intel\textsuperscript{\textregistered}
           Xeon\textsuperscript{\textregistered} X5550 CPU.
           The use of oblique cleaving planes and fine supersampling reduced sub-segment median voxel range relative to
           the median.
           \label{tab:CompoundAndNestedSubseg96ths}}
\end{table*}

A comparison of voxel counts and runtime for compound and nested segmentation is summarized in
\cref{tab:CompoundAndNestedSubseg96ths}. Using compound segmentation without supersampling or oblique cleaving planes
led to unusable data; for each sub-segment, at least one patient had a vanishingly small sub-segment encompassing zero
voxels. Using fine supersampling and oblique planes, sub-segment voxel counts had a mean of $6244.5$ voxels within each
sub-segment (std.\ dev.\ $=4417.6.0$; std.\ dev.\ of the mean $=96.1$; median $=5558.0$). The median number of voxels in
each sub-segment spanned $91.5-13829$, encompassing two orders of magnitude. Direct comparison of the voxel count
distributions within sub-segments via Kolmogorov-Smirnov tests showed the null hypothesis failed to be rejected and
distributions were found to differ significantly in $2435$ of $4560$ ($53.3\%$) unique test cases (\ie $p < 0.05$ in
$2435$ cases).

Nested segmentation was markedly different. Using nested segmentation \emph{without} supersampling or oblique planes,
sub-segment voxel counts had a mean of $27.2$ voxels within each sub-segment (std.\ dev.\ $=12.3$; std.\ dev.\ of the
mean $=0.15$; median $=24.0$). The median number of voxels in each sub-segment spanned $20-30$. Direct comparison of the
voxel count distributions within sub-segments via Kolmogorov-Smirnov tests showed the null hypothesis failed to be
rejected and distributions were found to differ significantly in $258$ of $4560$ ($5.7\%$) unique test cases (\ie $p <
0.05$ in $258$ cases). Applying the oblique planes method yielded a mean sub-segment voxel count of $19.4$ voxels within
each sub-segment (std.\ dev.\ $=9.0$; std.\ dev.\ of the mean $=0.04$; median $=19.0$). The median number of voxels in
each sub-segment spanned $18-20$. Direct comparison of the voxel count distributions within sub-segments yielded
significance in a single case out of $4560$ ($0.02\%$; \ie $p < 0.05$ for the Kolmogorov-Smirnov test in one case).
Applying supersampling with axis-aligned planes yielded a mean sub-segment voxel count of $4199.3$ voxels within each
sub-segment (std.\ dev.\ $=2139.3$; std.\ dev.\ of the mean $=9.7$; median = $4103.0$). The median number of voxels in
each sub-segment spanned $4027.5-4200$. No voxel count distributions were significantly distinct according to the
Kolmogorov-Smirnov test (\ie $p < 0.05$ in \emph{zero} of $4560$ tests).

Using both oblique planes and supersampling improved fairness of nested segmentation even more, though either oblique
planes or supersampling alone were sufficient for most purposes. The mean sub-segment voxel count was $4270.2$ voxels
within each sub-segment (std.\ dev.\ $=2004.8$; std.\ dev.\ of the mean $=9.1$; median $=4246.0$). The median number of
voxels in each sub-segment spanned $4212-4287$. Direct comparison of the voxel count distributions again found no
significantly distinct distributions (\ie $p < 0.05$ in \emph{zero} of $4560$ tests).

\section*{Discussion}

Planar segmentation can be accomplished using a variety of existing tools, \eg via Boolean structure combination
\citep{deasy2003cerr,pinter2012slicerrt}, conversion of \acp{ROI} to polygon surface meshes
and computing the intersection \citep{cgal49user,cgal49polygon} or via tessellation \citep{boots1999spatial}, 
or directly on \ac{ROI} contours via bisection \citep{clark2014automated}. Whatever the
method, sub-segments are effectively specified by the fractional volume between mutually orthogonal pairs of cleaving
planes. It may then seem intuitive that sub-segments with the same fractional volume between bounding planes, but at
different positions in the \ac{ROI}, would have the same volume. This intuition is valid for nested segmentation, but
not for compound segmentation. Compound segmentation only generates fair sub-segments when the \ac{ROI} is rectangular
and faces are aligned with the cleaving planes, and thus may lead to erroneous conclusions if used for sub-segment
comparison. A number of articles investigating the link between patient outcomes and radiotherapy dose to parotid
sub-volumes have recently emerged \citep{miah2016recovery,xiao2016split,clark2015regional} and use of compound
segmentation has been reported in the literature \citep{vanluijk2015sparing}. The aim of this study was to demonstrate
that nested segmentation is fairer than compound segmentation, and should be preferred for analyses involving
sub-segment comparison.

By analytically solving $\mathbb{R}^{2}$ and $\mathbb{R}^{3}$ analogues, we showed that compound method sub-segments
have intrinsically non-uniform area/volume. 
In $\mathbb{R}^{2}$, compound method centre sub-segment area differed from that of adjacent sub-segments by nearly a
\emph{third} of the fair area. The problem grew worse in $\mathbb{R}^{3}$ with the difference assuming more than
\emph{half} the fair volume. Nested method sub-segments were fair in both cases.

The successful segmentation of clinical \acp{ROI} into thirds indicates bisection is appropriate for locating cleaving
planes despite being unable to fairly partition due to discrete nature of contours along the axial direction. Compound
segmentation into $18$\textsuperscript{ths} was not fair. Distribution skewness and distinctness tests imply that the
parotid was not fairly partitioned into sub-segments of equivalent volume. At the same time, the lack of correlation
between sub-segment mean voxel count and relative position indicates the bisection approach is not systematically
biasing results and that sub-segment volumes appear to be comparable \emph{on average}. Nested segmentation, in
comparison, was fair. Distribution distinctness test results were substantially improved compared with compound
segmentation ($2$ \vs $124$ of $153$ tests found distinct distributions). The low number of distinct distributions
($1.3\%$) was comparable with the bisection tolerance ($1\%$) and therefore represents an acceptable deviation.
Performance on the Kolmogorov-Smirnov test is notable because the transverse cleave generally can not achieve fair
cleaving. The transverse cleave was performed first, and it is apparent that subsequent cleaves are fairer than 
those of compound segmentation.

The distinction between compound and nested segmentation was embiggened by segmentation into $96$\textsuperscript{ths}.
Some peripheral compound method sub-segments with vanishingly small volumes -- even when oblique planes and intensive
supersampling were employed. Nested method sub-segments were not quite fair when oblique planes and supersampling were
abstained from ($5.7\%$ of Kolmogorov-Smirnov tests were significant), but this was corrected when oblique planes,
supersampling, or both were employed ($0.02\%$ or less in all cases). The normalized range of voxels contained within a
sub-segment dropped when using oblique planes or supersampling, indicating sub-segment volumes became fairer when either
were employed. Compared to compound segmentation, nested segmentation produced normalized ranges that were two orders of
magnitude smaller. Additionally, \ac{QCD} differed by 1-2 orders of magnitude depending whether oblique planes were
used, suggesting nested method sub-segments were substantially less disperse, and thus more uniform, than compound
method sub-segments. These observation support the claim that nested segmentation is resilient to partitioning errors.
When a fair cleave could not be located, \ie due to discrete nature of contours along the axial direction, child
sub-segments were made as fairly as possible (\ie sub-segments equally shared the remaining volume with sibling
sub-segments). The increase of dispersion noted in compound method sub-segments when oblique planes were used supports
the claim that compound segmentation intrinsically can not fairly partition \acp{ROI}.

Nested segmentation was not only fairer than compound segmentation, but it was also faster. Sub-segments have planar
edges/faces that can be described with few vertices. In nested segmentation, recursive segmentation need only process a
simplified geometry for each planar edge/face. The full \ac{ROI} is processed once; afterwards, each additional
segmentation is continually reduced by the increasing number of planar edges. Compound segmentation, however, must
continually re-process the full \ac{ROI}. The exact speed-up depends on \ac{ROI} geometry, contour sampling density, and
nesting depth.

One downside of nested segmentation is that the shape of sub-segments depends on the order of cleaves, resulting
in shape asymmetries. 
A perfectly symmetrical segmentation method may be possible, but would most likely require iteration or back-tracking and
re-processing whole \acp{ROI} for each sub-segment. In contrast, nested segmentation requires neither back-tracking nor
re-processing geometry. Nested segmentation is directly applicable to organs where anatomical structure is ignorable or
\emph{a priori} unknown. It can also be employed within larger anatomical groupings, such as within lobes or cavities
(\eg liver, lung), and can make use of oriented cleaving planes or shuffled cleaving orders that align with local
anatomy (\eg muscle tissues, vessels, ducts). The use of planar segmentation combined with (iterated) bisection is a
flexible paradigm that enables the use of individual $\mathbb{R}^{2}$ contours, raster grids, disconnected collections
of contours, contours with holes, and volumetric surface manifolds, and would therefore be suitable addition to software
packages that can potentially operate on any such primitives (\eg \cite{pinter2012slicerrt,clark2014automated}).

Oblique cleaving planes addressed the issue of \ac{ROI} segment and voxel grid collinearity, but can result in awkward
plane orientations in some cases. There is an optimal cleave plane orientation that can be determined exactly when
sub-segment extents are known\footnote{The optimal angle is found for some special cases in $\mathbb{R}^{2}$ in a
supplementary document.}. This orientation maximizes the minimum spacing between voxel distances to the plane, ensuring
small changes in the plane position results in the smallest possible number of voxels crossing the plane at one time
(\eg minimizing spatial resonances). Unfortunately, even estimation is difficult and costly
\citep{fraser1965survey,dem1974introduction} so throughout this work a cyclic rotation of $22.5^{\circ}$ between cardinal
axes defined by the Cartesian dose grid was assumed. Supersampling is also useful for improving sub-segment fairness,
though it can not itself help the collinearity issue if planes are axes-aligned. However, when oblique planes and
supersampling are combined, supersampling will reduce the amount of obliquity needed, which can assist in adapting to
underlying anatomy. It will also result in sufficiently fair sub-segments if supersampling can be performed to an
arbitrary level, though it is computationally difficult and questionable to supersample too finely. Oblique planes were
more computationally efficient than supersampling, but application of either method independently for nested
segmentation into $96$\textsuperscript{ths} resulted in small median voxel ranges and acceptably indistinct
distributions (\ie $>99.9\%$ with $p>0.05$).

\subsection*{Conclusions}

Nested segmentation was found to be superior to compound segmentation when sub-segment volume consistency is needed.

\subsection*{Acknowledgments}

This work was partially funded by a University of British Columbia Four-Year Doctoral Fellowship and by a fellowship
from the Walter C.\ Sumner Foundation. The authors declare no conflicts of interest.

\newpage
\nolinenumbers
\bibliography{Bibliography}{}
\bibliographystyle{apalike}

\end{document}